\documentclass[fleqn,usenatbib]{mnras}

\usepackage{newtxtext,newtxmath}

\usepackage[utf8]{inputenc}
\usepackage[T1]{fontenc}
\usepackage{ae,aecompl}

\usepackage{graphicx}	
\usepackage{amsmath}	
\usepackage{amssymb}	

\usepackage{cprotect}
\usepackage[version=4]{mhchem}

\usepackage{listings}
\usepackage[normalem]{ulem}

\title[Carbon stars as standard candles]{Carbon stars as standard candles: I. The luminosity function of carbon stars in the Magellanic Clouds}

\author[Ripoche et al.]{
Paul Ripoche,$^{1,2}$\thanks{E-mail: pripoche@phas.ubc.ca}
Jeremy Heyl,$^{1}$
Javiera Parada,$^{1}$
Harvey Richer$^{1}$
\\
$^{1}$Department of Physics and Astronomy, University of British Columbia, Vancouver, British Columbia V6T 1Z1, Canada\\
$^{2}$École normale supérieure Paris-Saclay, 94235 Cachan Cedex, France\\
}

\date{Accepted XXX. Received YYY; in original form ZZZ}

\pubyear{2020}

\begin{document}
\label{firstpage}
\pagerange{\pageref{firstpage}--\pageref{lastpage}}
\maketitle

\begin{abstract}
Our goal in this paper is to derive a carbon-star luminosity function that will eventually be used to determine distances to galaxies at $50$-$60$ Mpc and hence yield a value of the Hubble constant. Cool N-type carbon stars exhibit redder near-infrared colours than oxygen-rich stars. Using Two Micron All Sky Survey near-infrared photometry and the \textit{Gaia} Data Release 2, we identify carbon stars in the Magellanic Clouds (MC) and the Milky Way (MW). Carbon stars in the MC appear as a distinct horizontal feature in the near-infrared ($(J-K_s)_0$, $M_J$) colour-magnitude diagram. We build a colour selection ($1.4 < (J-K_s)_0 < 2$) and derive the luminosity function of the colour-selected carbon stars. We find the median absolute magnitude and the dispersion, in the J band, for the Large Magellanic Cloud and the Small Magellanic Cloud to be, respectively, ($\bar{M_J} = -6.284~\pm~0.004$, $\sigma = 0.352~\pm~0.005$) and ($\bar{M_J} = -6.160~\pm~0.015$, $\sigma = 0.365~\pm~0.014$). The difference between the MC may be explained by the lower metallicity of the Small Magellanic Cloud, but in any case it provides limits on the type of galaxy whose distance can be determined with this technique. To account for metallicity effects, we developed a composite magnitude, named C, for which the error-weighted mean C magnitude of the MC are equal. Thanks to the next generation of telescopes (\textit{JWST}, \textit{ELT}, \textit{TMT}), carbon stars could be detected in MC-type galaxies at distances out to $50$-$60$ Mpc. The final goal is to eventually try and improve the measurement of the Hubble constant while exploring the current tensions related to its value.
\end{abstract}




\section{Introduction}
\label{section:introduction}

Measuring distances has been one of the most crucial, fascinating and challenging goals in astronomy for centuries. Not only does it enable astronomers to probe the scale of the Universe, it is also key in understanding the physical nature of astronomical objects, which eventually opens up whole new fields of astrophysical research.

Carbon stars were discovered by \citet{1868MNRAS..28..196S} because of a characteristic feature in their spectra: the Swan bands, which are representative of radical diatomic carbon \ce{C2}. Carbon stars are usually luminous red giant stars. Bright carbon stars are located on the asymptotic giant branch (AGB), in the Hertzsprung-Russell diagram. Carbon stars originate from main-sequence stars with masses in the range $1-10~\text{M}_{\odot}$ \citep{1991AJ....102..289B, 1999A&A...344..123M}. During the third dredge-up, heavy elements (\ce{C}, \ce{O}) are brought to the surface from the stellar interior \citep{1984PhR...105..329I, 1999ARA&A..37..239B}. In normal (oxygen) stars $\ce{O}>\ce{C}$ (in terms of number of atoms), nearly all of the carbon in the atmosphere is trapped in the stable compound \ce{C}\ce{O}. However, if $\ce{C}/\ce{O} > 1$, carbon compounds can appear in the stellar atmosphere of AGB stars, forming carbon stars.

In this work, we focus on cool N-type carbon stars; such carbon stars have been known for a long time to show redder near-infrared colours than M-type stars (oxygen-rich stars). Thus, cool carbon stars occupy a specific region of near-infrared colour-magnitude diagrams \citep{1979ApJ...230..724R, 1981ApJ...249..481C, 1990AJ.....99..784H}. There is also evidence of similar carbon-star luminosities between the Magellanic Clouds \citep{1976ApJ...203L..81C, 1980ApJ...242..938B, 1981ASSL...88..153R, 1981ApJ...243..744R, 1982ApJS...48..161A, 1983ApJ...275...84F, 1992MNRAS.259....6F, 1993MNRAS.260..103D}. \citet{2002AJ....123.3428D} obtained luminosity functions for a small sample of carbon stars in the Magellanic Clouds using a near-infrared technique and a colour selection similar to the one used in this paper. Since carbon stars are very luminous in the infrared and have a characteristic spectrum, the possibility exists that they can be used as physical distance indicators \citep{1979ApJ...230..724R, 1995AJ....109.2480B, 1999ApJ...524L.111V, 2005AJ....129..729R}, i.e. standard candles. \citet{1985ApJ...298..240R} used the carbon stars in NGC 300 to measure its distance modulus. The value they found agreed with the value derived from Cepheids; this was the first time carbon stars were used as distance indicators.

Some other standard candles used by astronomers are the tip of the red-giant branch \citep{2017ApJ...836...74J, 2018ApJ...858...12H, 2018ApJ...852...60J, 2019ApJ...882...34F, 2019ApJ...886...61Y}, Cepheids \citep{2019MNRAS.490.4975R, 2019ApJ...876...85R} and Type Ia supernovae \citep{2018ApJ...853..126R}. However, Type Ia supernovae have been known to show properties that actually differ with distance \citep{2015ApJ...803...20M, 2020MNRAS.491.5991F}. In addition, the Cepheid period-luminosity relation has a metallicity sensitivity that increases towards lower-metallicity regimes \citep{2017ApJ...842..116W, 2018A&A...620A..99G}. Finally, the tip-of-the-red-giant-branch (TRGB) stars are typically less bright than carbon stars by about 1 magnitude, and the TRGB luminosity depends on stellar age and metallicity \citep{2008ApJ...689..721M, 2019ApJ...880...63M}.

Our goal in this paper is to determine an accurate selection criteria and derive a carbon-star luminosity function (CSLF) in order to eventually use these stars as standard candles. We first develop an accurate selection criteria to identify these stars in distant galaxies. We then derive the CSLF in both of the Magellanic Clouds (MC), using the Two Micron All Sky Survey All-Sky Point Source Catalogue (2MASS PSC). 
We then attempt to reproduce the same process in our Galaxy, the Milky Way (MW), using the \textit{Gaia} Data Release 2 (\textit{Gaia} DR2), along with the 2MASS catalogue.

\section{Carbon stars in the Large Magellanic Cloud}
\label{chapter:carbonstarsLMC}
\subsection{Photometry of stars in the LMC}
\label{section:photometrystarsLMC}

\subsubsection{2MASS query and removing galactic foreground stars}
We used the 2MASS PSC in order to get a near-infrared catalogue of stars in the LMC. 2MASS is an astronomical survey that has uniformly scanned the entire sky in three near-infrared bands: $J$ ($1.235$ $\umu$m), $H$ ($1.662$ $\umu$m) and $K_s$ ($259$ $\umu$m). 2MASS used two highly-automated $1.3$ m telescopes, one on Mt. Hopkins, AZ, the other in Chile at CTIO. 

We queried the 2MASS PSC using the following polygon query: \texttt{polygon(65 -76, 65 -62, 100 -62, 100 -76)}. Each input vertex of the polygon is a J2000 R.A. and Dec pair, in decimal degrees. From this query we obtained a full catalogue of stars in the LMC with photometry in the 2MASS bands.

\begin{figure}
\includegraphics[width=\columnwidth]{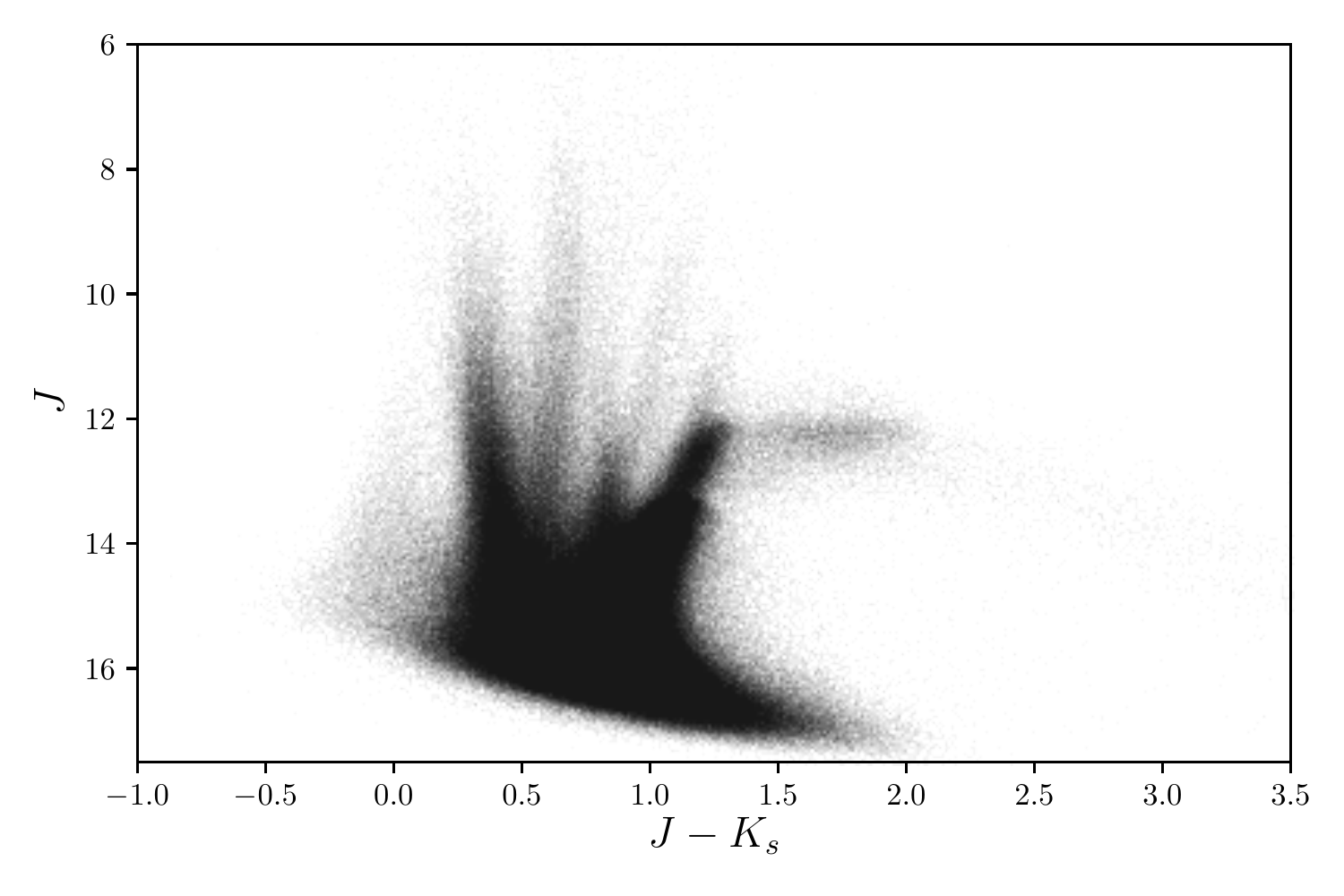}
    \caption{Colour-magnitude diagram of the Large Magellanic Cloud, in the near infrared 2MASS bands. We plotted the apparent magnitude as retrieved from the 2MASS PSC.}
    \label{fig:CMDapparLMCraw}
\end{figure}

\begin{figure}
\includegraphics[width=\columnwidth]{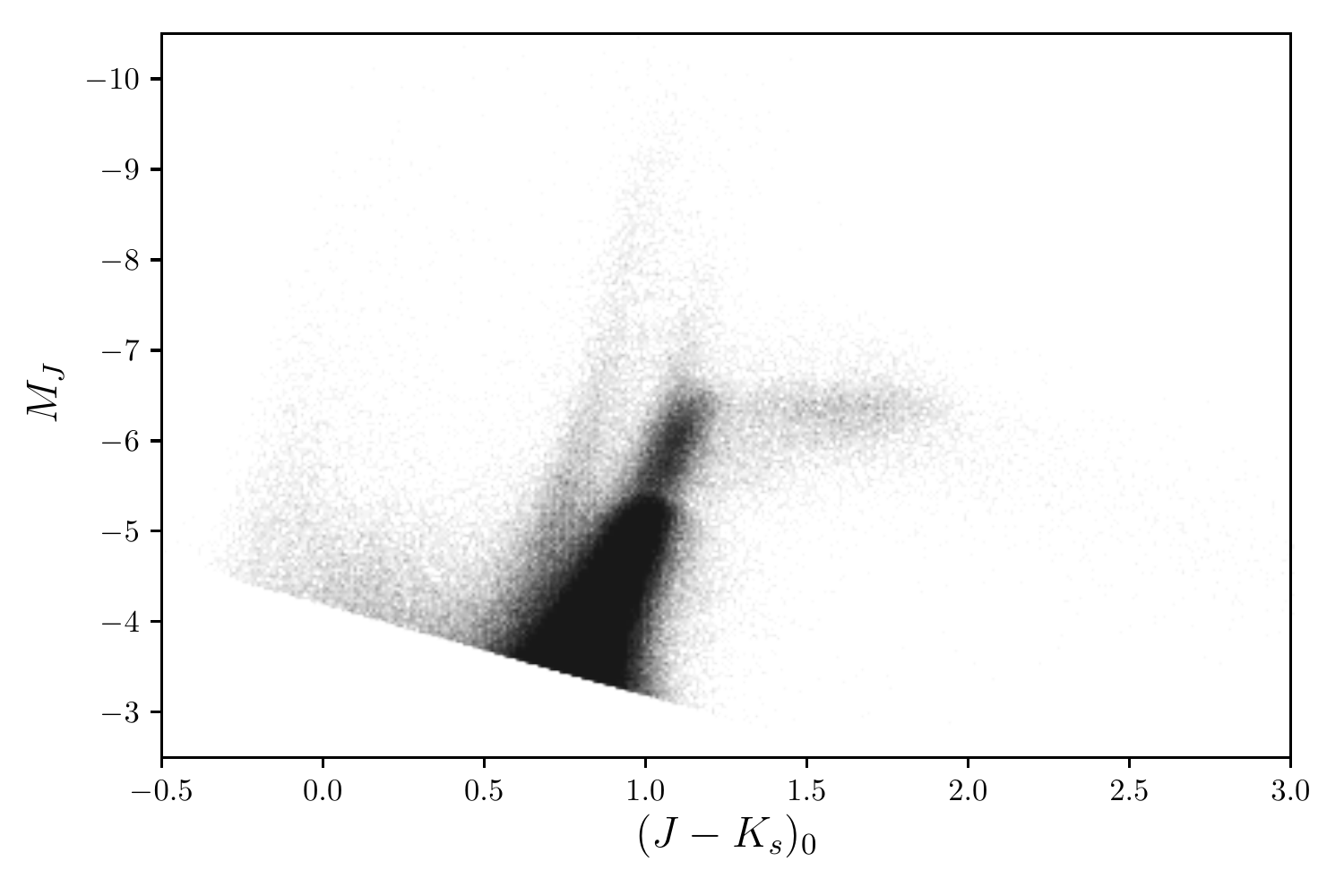}
    \caption{Colour-magnitude diagram of the Large Magellanic Cloud, in the near infrared 2MASS bands. We plotted the absolute magnitude using the 2MASS data from our filtered catalogue, corrected for reddening.}
    \label{fig:CMDLMC2MASSfiltered}
\end{figure}

Using the 2MASS photometry we plotted the colour-magnitude diagram (CMD), in apparent magnitude of this full catalogue of stars in the LMC (see Fig.~\ref{fig:CMDapparLMCraw}). We note that there is a strong galactic foreground represented by the first two ``fingers'' in the CMD. Moroever, objects below $J = 16$ have a poor signal-to-noise ratio \citep{2006AJ....131.1163S}. The completeness of the 2MASS PSC is described in table~\ref{table:completeness2MASS}. 
\begin{table}
\caption{$99$\% completeness limit magnitude of the 2MASS PSC \citep{2006AJ....131.1163S}}
\centering
\begin{tabular}{c c c}
\hline\hline
$J$ & $H$ & $K_s$ \\ [0.5ex]
\hline
15.8 & 15.1 & 14.3 \\[1ex]
\hline
\hline
\end{tabular}
\label{table:completeness2MASS}
\end{table}
We filter our catalogue by selecting the stars brighter than the 2MASS completeness limit, in each band. The horizontal feature of carbon stars is located at about $J=12$ in the LMC (see Fig.~\ref{fig:CMDapparLMCraw}), therefore we should expect 100\% completeness for our selection of carbon stars. We define:
\begin{itemize}
    \item[-] $G \equiv \verb!phot_g_mean_mag!$
    \item[-] $G_{BP} \equiv \verb!phot_bp_mean_mag!$
    \item[-] $G_{RP} \equiv \verb!phot_rp_mean_mag!$
    \item[-] $\chi^2 \equiv \verb!astrometric_chi2_al!$
    \item[-] $\nu \equiv \verb!astrometric_n_good_obs_al! - 5$
    \item[-] $E \equiv \verb!phot_bp_rp_excess_factor!$,
\end{itemize}
where the name of the parameters are the same as in the \textit{Gaia} documentation. We cross-correlated our 2MASS data with the \textit{Gaia} DR2, keeping only the sources with a five-parameter solution \citep{2018A&A...616A...2L}:
\begin{equation} 
\left.
    \begin{array}{l}
        \verb!phot_g_mean_mag! < 21 \\
        \verb!visibility_periods_used! \geq 6 \\
        \verb!astrometric_sigma5d_max! < 1.2 \times \gamma(G)
    \end{array}
\right\},
\label{five_parameter}
\end{equation}
where $\gamma (G) = \text{max}[1, 10^{0.2(G-18)}]$. We also removed spurious solutions \citep{2018A&A...616A...2L}:
\begin{equation} 
\left.
    \begin{array}{l}
        u < 1.2 \times \text{max}\left\{1, \exp[-0.2(G-19.5)]\right\} \\
        E > 1 + 0.015(G_{BP}-G_{RP})^2 \\
        E < 1.3 + 0.06(G_{BP}-G_{RP})^2
    \end{array}
\right\},
\label{equation:spurious}
\end{equation}
where $u = \sqrt{\chi^2/\nu}$. Finally we reduced galactic foreground using:
\begin{equation} 
\left.
    \begin{array}{l}
    (\verb!pmra! - 1.850)^2 + (\verb!pmdec!- 0.234)^2 < 1 \\
     (\verb!pmra!/\verb!pmra_error!)^2 + (\verb!pmdec!/\verb!pmdec_error!)^2 > 25 \\
    \verb!astrometric_matched_observations! \geq 8 \\
    |\verb!parallax_over_error!| < 5 \\
    \verb!parallax_error! < 1 \\
    |\sin(\verb!b!)| > 0.1
    \end{array}
\right\}.
\label{equation:galactic_foreground}
\end{equation}

The first condition of Eq.~(\ref{equation:galactic_foreground}) is a selection of stars obtained by comparing their proper motions to those of the LMC \citep{2018A&A...616A...2L}; we used  the centre-of-mass proper motions of the LMC described in \citet{2018A&A...616A..12G}. The following conditions in Eq.~(\ref{equation:galactic_foreground}) reduce the contamination by galactic foreground \citep{2018A&A...616A...2L}. The number of retrieved stars after each filtering is summarised in table~\ref{table:completeness2MASSnumber}. We now have a catalogue of stars in the LMC, on which we can work to identify the carbon stars. 

\begin{table}
\caption{Number of carbon stars in our LMC catalogue at each step of the reducing process.}
\centering
\begin{tabular}{l c}
\hline\hline
2MASS query & 1,397,958 \\ [0.5ex]
\hline
2MASS query complete ($J$, $H$, $K_s$) & 563,357 \\[0.5ex]
\hline
stars with a five-parameter solution & 547,335 \\[0.5ex]
\hline
discarding spurious solutions & 492,904 \\[0.5ex]
\hline
reducing galactic foreground & 310,187 \\[1ex]
\hline
\hline
\end{tabular}
\label{table:completeness2MASSnumber}
\end{table}

\subsubsection{Absolute magnitude of LMC stars}
After having removed galactic foreground stars, we now have to transform apparent magnitudes to absolute magnitudes. The first step is to include the mean distance modulus to the LMC, and then we need to correct for extinction (magnitude) and reddening (colour).

\citet{2019Natur.567..200P} determined a geometrical distance to the LMC based on 20 eclipsing binary systems that is accurate to $1 \%$. They found a mean distance modulus: $\mu_\text{LMC} = 18.477 \pm 0.004$ (statistical) $\pm~0.026$ (systematic). This is the value we will use to get the absolute magnitude of the stars in our catalogue.

\citet{2020ApJ...889..179G} obtained new reddening maps of the MC based on colour measurements of the red clump. $E(B-V)$ values (the colour excess in the $B$ and $V$ bands) were obtained by calculating the difference of the observed and intrinsic colour of the red clump in the LMC and the SMC. They found a mean value of the reddening for the LMC of $E(B-V)_{\text{LMC}} = 0.127 \pm 0.013$. We adopt this value and the absolute magnitude in 2MASS bands is given by:
\begin{equation}
    M_x = x - \mu_\text{LMC} - \frac{k_x}{k_B-1}E(B-V)_{\text{LMC}},
    \label{equation:absolutemagnitude}
\end{equation}

where $x \in \{J, H, K_s\}$, $k_x = A_x/A_V$ and $k_B = A_B/A_V$; $A$ is the extinction in a given band. The $k$ coefficients are given in table \ref{table:reddeningMC}. Similarly the intrinsic colours are given by:
\begin{equation}
    (J-K_s)_0= (J-K_s) - \frac{k_J - k_{K_s}}{k_B-1}E(B-V)_{\text{LMC}}
\end{equation}
\begin{equation}
    (J-H)_0 = (J-H) - \frac{k_J - k_H}{k_B-1}E(B-V)_{\text{LMC}}
\end{equation}
\begin{equation}
    (H-K_s)_0 = (H-K_s) - \frac{k_H - k_{K_s}}{k_B-1}E(B-V)_{\text{LMC}},
    \label{equation:colours}
\end{equation}
where $J-K_s$, $J-H$ and $H-K_s$ are the observed colours from the 2MASS PSC.

\begin{table}
\caption{$k_\lambda$ coefficients in 2MASS bands for the MC \citep{2003ApJ...594..279G}}
\centering
\begin{tabular}{c c c c c}
\hline\hline
& $k_J$ & $k_H$ & $k_{K_s}$ & $k_B$ \\ [0.5ex]
\hline
\textbf{LMC} & 0.257 & 0.186 & 0.030 & 1.293 \\[1ex]
\textbf{(average)} & $\pm~0.013$ & $\pm~0.020$ & $\pm~0.003$ & $\pm~0.113$ \\[1ex]
\hline
\textbf{SMC} & 0.131 & 0.169 & 0.016 & 1.374 \\[1ex]
\textbf{(bar)} & $\pm~0.013$ & $\pm~0.020$ & $\pm~0.003$ & $\pm~0.127$ \\[1ex]
\hline
\hline
\end{tabular}
\label{table:reddeningMC}
\end{table}

The filtered and reddening-corrected colour-magnitude diagram of the LMC in absolute magnitude is plotted in Fig.~\ref{fig:CMDLMC2MASSfiltered}. Carbon stars appear as a horizontal feature in the CMD of the LMC (see Fig.~\ref{fig:CMDLMC2MASSfiltered}), at about $M_J \approx -6$. This ``red tail'' of carbon stars is in agreement with simulations of LMC stars by \citet{2003A&A...403..225M, 2008A&A...482..883M}.

We can now distinguish more precisely the different types of stars in the CMD of the LMC. The main ``finger'' is the red giant branch (RGB), ended by the TRGB. On top of the TRGB is the AGB (oxygen-rich stars). To the left of the RGB are the red supergiants (the faint ``finger''), and to the left of that, the main-sequence stars at $(J-K_s)_0 \approx 0$. Finally the distinct horizontal feature contains the carbon stars.

\subsection{Identifying carbon stars in the LMC}
\label{section:colorselectionLMC}

\subsubsection{Spectroscopic catalogue}
In order to identify the nature of the stars that appear in this horizontal feature, we used the spectroscopic catalogue of carbon stars in the LMC by \citet{2001A&A...369..932K}; a catalogue of 7,760 carbon stars. The stars were identified during a systematic survey of objective-prism plates taken with the UK $1.2$ m Schmidt Telescope.

\begin{figure}
\includegraphics[width=\columnwidth]{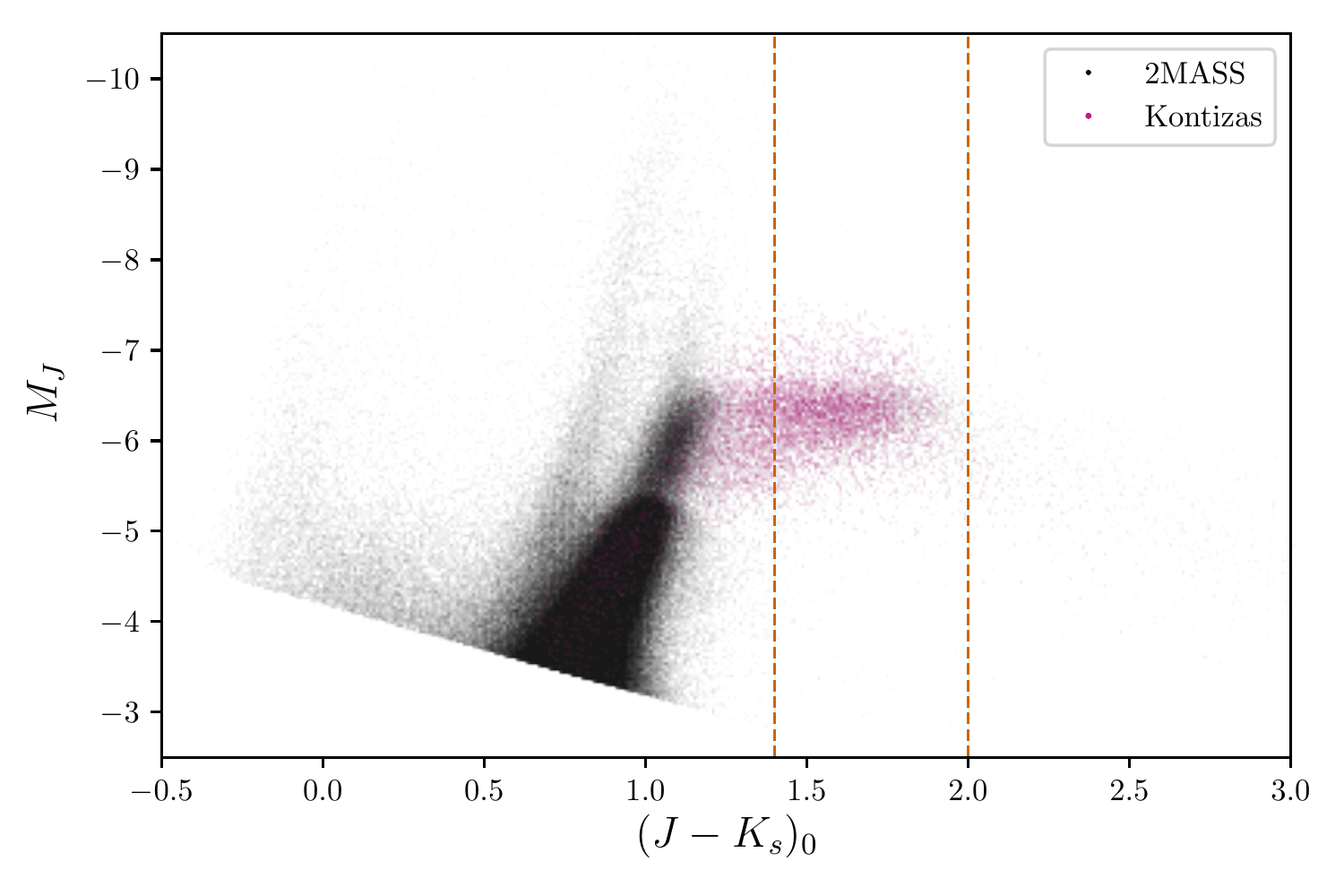}
    \caption{Colour-magnitude diagram of the Large Magellanic Cloud. In black are the stars from our filtered and reddening-corrected catalogue; in magenta are the stars from \citet{2001A&A...369..932K}. The orange dashed lines represent the limits of the colour selection.}
    \label{fig:CMDLMCKontizas2MASS}
\end{figure}

We cross correlated this catalogue with the 2MASS PSC in order to obtain the near-infrared photometry of the carbon stars. As we did in section~\ref{section:photometrystarsLMC}, we filtered it; this reduced the catalogue to $7,279$ stars. In Fig.~\ref{fig:CMDLMCKontizas2MASS} we over-plotted the two catalogues in order to compare them. We find in this CMD the horizontal feature that appears in Fig.~\ref{fig:CMDLMC2MASSfiltered}. The Kontizas catalogue seems to also contain warmer carbon stars (C-R or C-H types), see section~\ref{section:spectroscopic_MW} for more details on the different types of carbon stars.

\subsubsection{Colour selection}
The $(J-K_s)_0$ colour of bright carbon stars is known to be greater than $1.4$ \citep{2011A&A...530A...8W}. Indeed, carbon stars seem to appear well separated from oxygen-rich AGB stars in the colour-magnitude diagram (see Fig.~\ref{fig:CMDLMCKontizas2MASS}), for $(J-K_s)_0 > 1.4$.

The fact that carbon stars could be identified in the CMD of the LMC by only making a colour selection enables us to apply this method to objects with unknown distances. We therefore need to know if the stars that appear in these distinct features are actually carbon stars. 

Based on the horizontal feature of the CMD for the LMC, we build a colour selection such that: $1.4 < (J-K_s)_0 < 2$ (see Fig.~\ref{fig:CMDLMCKontizas2MASS}). This colour selection is in agreement with \citet{2001A&A...377..945C, 2003A&A...406...51C}.

Sixty-nine percent of the 2MASS stars in this colour range were matched with carbon stars from the Kontizas catalogue in the exact same colour range; this represents $98 \%$ of the stars from the Kontizas catalogue in this colour range. This means that most of the remaining $31 \%$ of stars are possible carbon stars that have not been identified yet using spectroscopy.

\subsection{Carbon-star luminosity function in the LMC}
\begin{figure}
\includegraphics[width=\columnwidth]{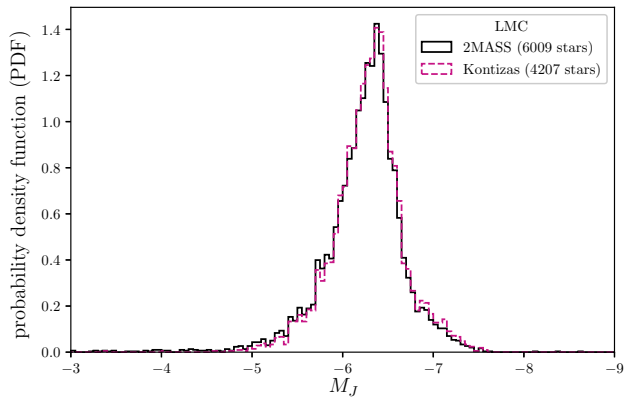}
    \caption{Carbon-star luminosity function in the Large Magellanic Cloud, for stars within the colour selection. In magenta are the stars from the Kontizas catalogue; in black are the stars from our 2MASS query.}
    \label{fig:luminosityfunctionKontizas2MASSLMC}
\end{figure}

Now that we are able to identify carbon stars using only the $(J-K_s)_0$ colour, we need to study the carbon star luminosity function in the LMC. A luminosity function gives the number of stars per absolute magnitude interval. We used the absolute magnitude in the J-band of 2MASS. We derived the CSLF of the LMC for all the stars within the following colour range (see the black histogram in Fig.~\ref{fig:luminosityfunctionKontizas2MASSLMC}): 
\begin{equation}
    1.4 < (J-K_s)_0 < 2.
\end{equation}
The luminosity function of the stars from our 2MASS query and of the ones from the Kontizas catalogue are plotted in Fig.~\ref{fig:luminosityfunctionKontizas2MASSLMC} for comparison purposes. 

In order to estimate the median absolute magnitude $\bar{M}_J$ and the intrinsic magnitude dispersion $\sigma$ of the carbon stars in the LMC, we use maximum likelihood statistics. The errors in $J$, reddening and distance are assumed to follow a normal distribution and to be uncorrelated. We also assume that for a sample of $n$ stars, the values $\{M_{J,i}; ...; M_{J,n}\}$ have a Gaussian distribution centered on $\bar{M}_J$. However, the luminosity function in Fig.~\ref{fig:luminosityfunctionKontizas2MASSLMC} contains carbon stars from about $M_J=-3$ to $M_J=-8.5$. Even though the number of these outliers is small, they can still affect the statistics of the distribution. Consequently, instead of using the standard deviation, which is affected by outliers in a data set, or even the more robust median absolute deviation, we use the estimator $Q_n$ \citep{rousseeuw}. $Q_n(M_J)$ is given by the $0.25$ quantile of the distances $ \{ \left|M_{J,i} - M_{J,j}\right| ; i < j \} $. We then numerically estimate $\bar{M}_J$ and $\sigma$ by minimizing the following function $L$, which is the logarithm of the inverse of the joint probability function of a Gaussian distribution:
\begin{equation}
\label{eq:maximumlikelihood}
    \begin{split}
        L =~& n \left\{Q_n^2\left(\frac{M_J-\bar{M}_J}{\left[2(\sigma_{M_J}^2+\sigma^2)\right]^{1/2}}\right) \right.\\
        & \left. +~med^2\left(\frac{M_J-\bar{M}_J}{\left[2(\sigma_{M_J}^2+\sigma^2)\right]^{1/2}}\right)\right\} \\
        & +~ \sum_{i=1}^{n} \frac{1}{2}\ln{(\sigma_{M_{J,i}}^2+\sigma^2)},
    \end{split}
\end{equation}
where $n$ is the number of stars in the colour window, the estimator $med$ returns the median value, and $\sigma_{M_J}$ is the error on $M_J$. 
We find for our 2MASS query:
\begin{equation} 
\left\{
    \begin{array}{l}
    \bar{M}_J = -6.284 \pm 0.004\\
    \sigma = 0.352 \pm 0.005,
    \end{array}
\right.
\end{equation}
and for the Kontizas catalogue:
\begin{equation} 
\left\{
    \begin{array}{l}
    \bar{M}_J = -6.306 \pm 0.006\\
    \sigma = 0.337 \pm 0.006.
    \end{array}
\right.
\end{equation}

To summarise, we have developed a purely photometric method, based on one near-infrared colour, $(J-K_s)$, to identify carbon stars in the LMC. Since our method depends only on colour (a parameter independent of distance) one can apply it to distant galaxies using apparent magnitudes.

\section{Carbon stars in the Small Magellanic Cloud}
\label{chapter:carbonstarsSMC}

We aim to reproduce our LMC results in the Small Magellanic Cloud, using the same methods.

\subsection{Photometry of stars in the SMC}
\label{section:photometrystarsSMC}

\subsubsection{2MASS query and removing galactic foreground stars}
As for the LMC, we used the 2MASS PSC in order to obtain a near-infrared catalogue of stars in the SMC. We queried the 2MASS PSC using the following polygon: \texttt{polygon(7.78 -74.15, 18.30 -74.15, 18.30 -71.43, 7.78 -71.43)}. Each input vertex of the polygon is a J2000 R.A. and Dec pair, in decimal degrees. From this query we obtained a full catalogue of the stars in the SMC, in the 2MASS bands.

\begin{figure}
\includegraphics[width=\columnwidth]{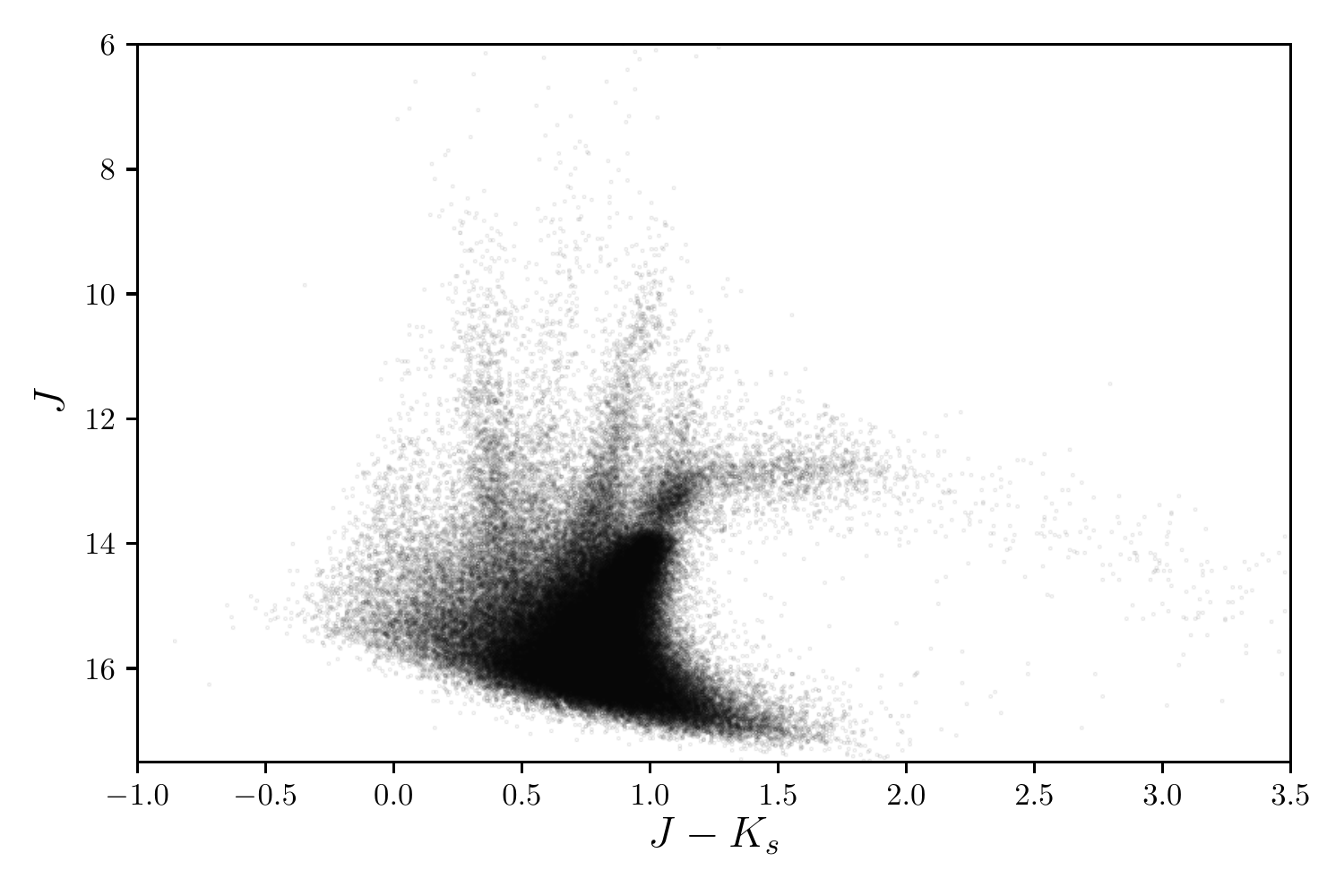}
\caption{Colour-magnitude diagram of the Small Magellanic Cloud in the near infrared 2MASS bands. We plotted the apparent magnitude as retrieved from the 2MASS PSC.}
\label{fig:CMDapparSMCraw}
\end{figure}

Using the SMC 2MASS photometry, we noticed the same strong galactic foreground that needs to be removed (see Fig.~\ref{fig:CMDapparSMCraw}). We reduced our SMC catalogue the same way as for the LMC (see section.~\ref{section:photometrystarsLMC}). We only change the first condition of Eq.~(\ref{equation:galactic_foreground}); we used the centre-of-mass proper motions of the SMC described in \citet{2018A&A...616A..12G}:
\begin{equation} 
    (\verb!pmra! - 0.797)^2 + (\verb!pmdec! + 1.220)^2 < 1
\end{equation}
We now have a catalogue of stars in the SMC in which we can identify the carbon stars. The number of retrieved stars after each filtering is summarised in table \ref{table:completeness2MASSnumberSMC}.

\begin{table}
\caption{Number of carbon stars in our SMC catalogue at each step of the reducing process.}
\centering
\begin{tabular}{l c}
\hline\hline
2MASS query & 128,252 \\ [0.5ex]
\hline
2MASS completeness & 45,249 \\[0.5ex]
\hline
stars with a five-parameter solution & 43,746 \\[0.5ex]
\hline
discarding spurious solutions & 40,100 \\[0.5ex]
\hline
reducing galactic foreground & 33,372 \\[1ex]
\hline
\hline
\end{tabular}
\label{table:completeness2MASSnumberSMC}
\end{table}

\subsubsection{Absolute magnitude of SMC stars}
After having removed galactic foreground stars, we need to transform apparent magnitude to absolute magnitude. The first step is to include the mean distance modulus to the SMC, and then we need to correct for reddening and extinction. 

Using \textit{Spitzer} observations of classical Cepheids, \citet{2016ApJ...816...49S} determined a distance to the SMC. They found a mean distance modulus: $\mu_\text{SMC} = 18.96\pm 0.01$ (statistical) $\pm~0.03$ (systematic). This the value we use for our catalogue of stars in the SMC. 

We again use $E(B-V)$ values calculated by \citet{2020ApJ...889..179G}. They found a mean value of the reddening for the SMC of $E(B-V)_{\text{SMC}} = 0.084 \pm 0.013$. The $k$ coefficients for the SMC are given in table \ref{table:reddeningMC}. The filtered and reddening-corrected colour-magnitude diagram of the SMC, in absolute magnitude, is plotted in Fig.~\ref{fig:CMDSMCRaimondo2MASS} (black stars).

\begin{figure}
\includegraphics[width=\columnwidth]{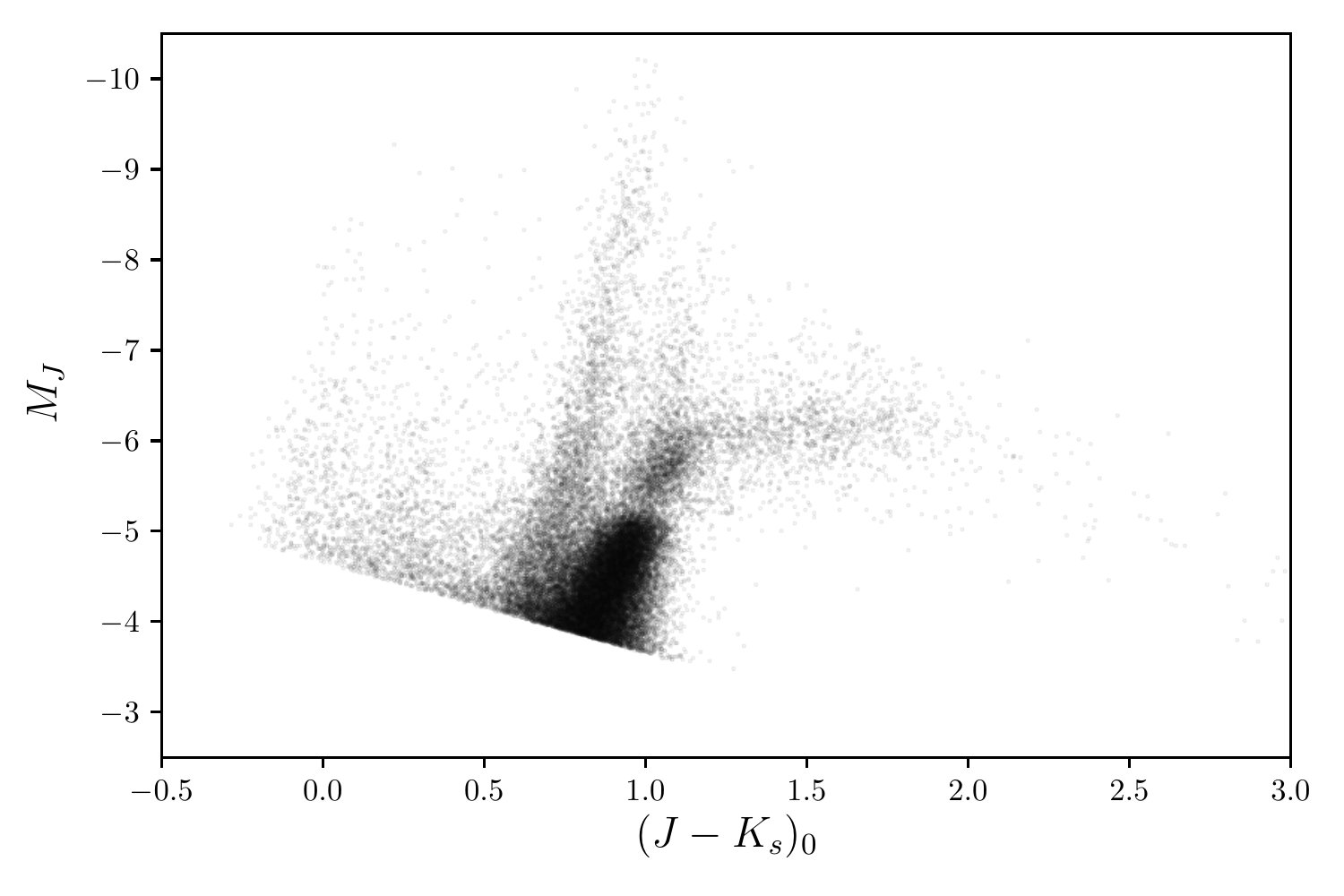}
\caption{Colour-magnitude diagram of the Small Magellanic Cloud, in the near infrared 2MASS bands. We plotted the absolute magnitude using of the data from our filtered catalogue, corrected for reddening.}
\label{fig:CMDSMC2MASSfiltered}
\end{figure}

As was the case for the LMC, we can now distinguish more precisely the different types of stars in the CMD of the SMC (see Fig.~\ref{fig:CMDSMC2MASSfiltered}); i.e. the RGB, ended by the TRGB, the AGB, the red super giants and the main-sequence stars. Finally, we find the same distinct horizontal feature that hosts carbon stars.

\subsection{Identifying carbon stars in the SMC}
\label{section:colorselectionSMC}

\subsubsection{Spectroscopic catalogue}
In order to identify the nature of the stars that appear in this horizontal feature, we used the spectroscopic catalogue of carbon stars in the SMC by \citet{2005A&A...438..521R}. This is a catalogue of $1079$ carbon stars from massive compact halo object (MACHO) observations. This sample includes confirmed carbon stars from the spectroscopic atlas by \citet{1993A&AS...97..603R}.

\begin{figure}
\includegraphics[width=\columnwidth]{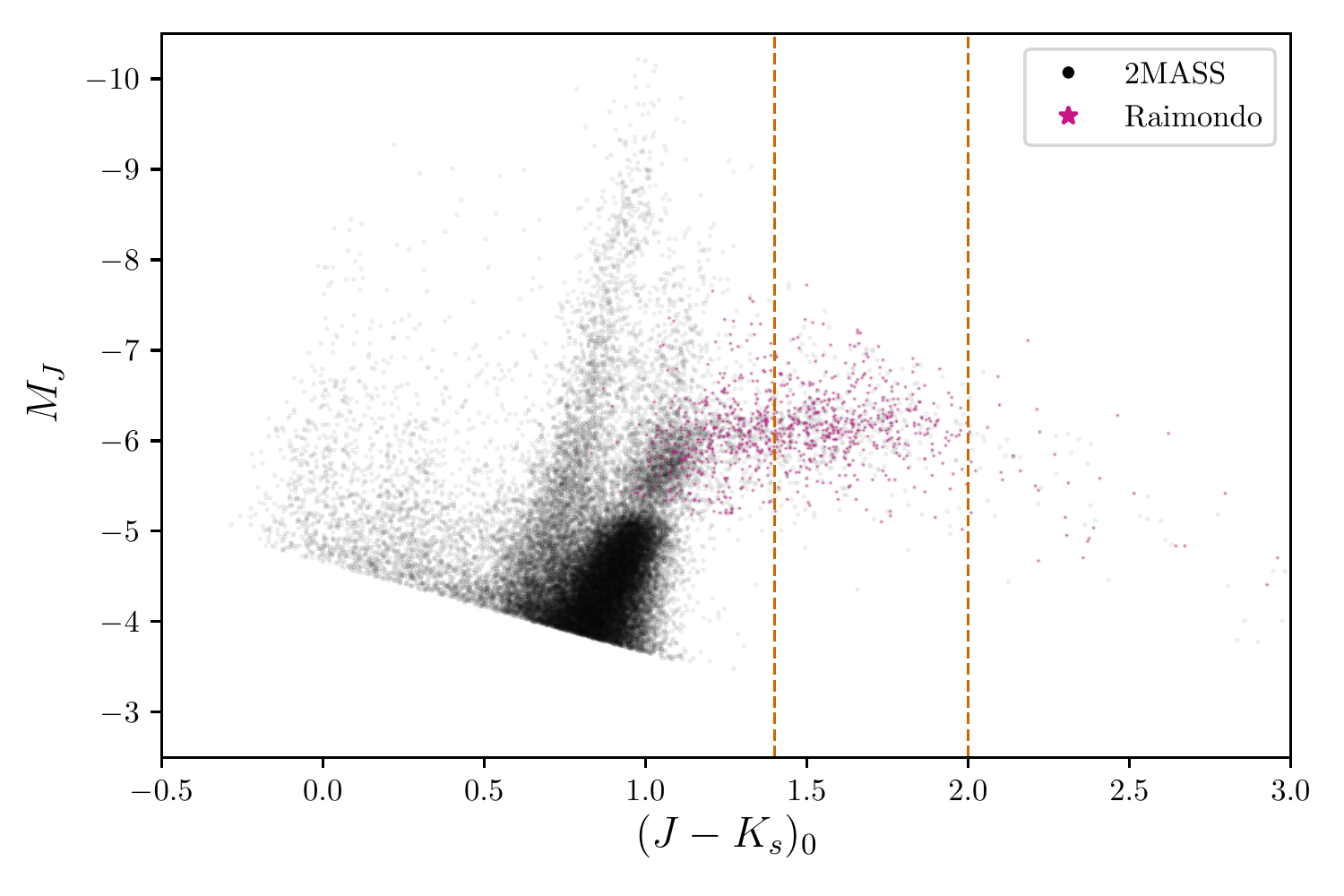}
\caption{Colour-magnitude diagram of the Small Magellanic Cloud. In black are the stars from our filtered-and-corrected for reddening catalogue; in magenta are the stars from \citet{2005A&A...438..521R}.}
\label{fig:CMDSMCRaimondo2MASS}
\end{figure}

We cross correlated this catalogue with the 2MASS PSC in order to obtain the near-infrared photometry of the carbon stars. As for the SMC 2MASS query, we filtered it to remove galactic foreground stars. During this process the catalogue was reduced to $920$ stars. We plot the CMD of these stars in Fig.~\ref{fig:CMDSMCRaimondo2MASS} (magenta stars). We find that it displays the horizontal feature of carbon stars.

\subsubsection{Colour selection}
As in the LMC, carbon stars seem to appear well separated from oxygen-rich AGB stars in the colour-magnitude diagram of the SMC for $(J-K_s)_0 > 1.4$.

Based on the horizontal feature in the CMD of the SMC, we build a similar colour selection such that: $1.4 < (J-K_s)_0 < 2$.

Fifty-five percent of the 2MASS stars in this colour range were matched with carbon stars from the Raimondo catalogue in the exact same colour range. This represents $100 \%$ of the stars from the Raimondo catalogue in this colour range. As for the LMC, most of the non-identified stars in the feature should be carbon stars.

\subsection{Carbon-star luminosity function in the SMC}
\begin{figure}
\includegraphics[width=\columnwidth]{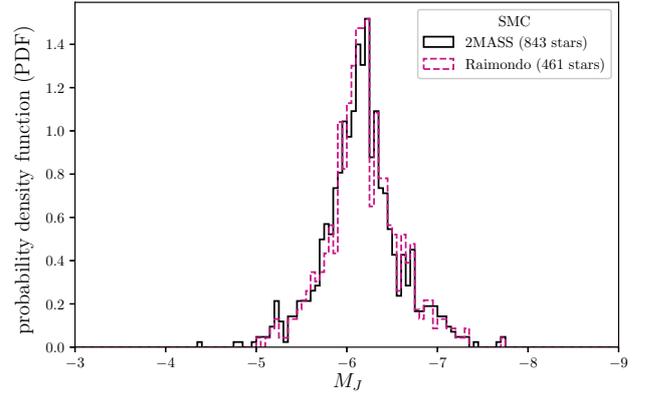}
\caption{Carbon-star luminosity function in the Small Magellanic Cloud, for stars within the colour selection. In magenta are the stars from the Raimondo catalogue; in black are the stars from our 2MASS query.}
\label{fig:luminosityfunctionRaimondo2MASSSMC}
\end{figure}

Now that we are able to identify carbon stars using only the $(J-K_s)_0$ colour, we study the carbon-star luminosity function in the SMC. We derived the J-band CSLF of the SMC for all the stars within the colour range (see the black histogram in Fig.~\ref{fig:luminosityfunctionRaimondo2MASSSMC}): 
\begin{equation}
    1.4 < (J-K_s)_0 < 2.
\end{equation}

The luminosity function of the stars from our 2MASS query and of the ones from the Raimondo catalogue are plotted in Fig.~\ref{fig:luminosityfunctionRaimondo2MASSSMC} for comparison purposes. 

In order to estimate the median absolute magnitude $\bar{M}_J$ and the intrinsic magnitude dispersion $\sigma$ of the carbon stars in the SMC, we use maximum likelihood statistics. We use the same assumptions as for the LMC and Eq.~\ref{eq:maximumlikelihood}. We obtain for our 2MASS query:
\begin{equation} 
\left\{
    \begin{array}{l}
    \bar{M}_J = -6.160 \pm 0.015\\
    \sigma = 0.365 \pm 0.014,
    \end{array}
\right.
\end{equation}
and for the Raimondo catalogue:
\begin{equation} 
\left\{
    \begin{array}{l}
    \bar{M}_J = -6.172 \pm 0.015\\
    \sigma = 0.349 \pm 0.018.
    \end{array}
\right.
\end{equation}

To conclude, by using the same method as for the LMC, we were able to identify carbon stars in another galaxy, the SMC. We also observe a very similar and well-defined CSLF in both galaxies.

\section{Carbon stars in the Milky Way}
\label{chapter:carbonstarsMW}

In what has transpired above, we used the LMC carbon stars as standard candles in order the establish a luminosity function that will eventually be used in distant galaxies. This required the distance to the LMC determined by some technique. A better approach might be to derive the carbon star luminosity function directly in the Milky Way Galaxy in order to avoid this step.

\subsection{Photometry}
\label{section:photometryMW}

If the distribution of carbon stars in the MW matches the one in the MC, we could use the MW CSLF in order to calibrate the distance modulus to the LMC. 
To explore this, we first use the catalogue of galactic carbon stars established by \citet{2001BaltA..10....1A}. To compare the CMD of these stars to the one in the LMC, we need the corresponding photometry in the 2MASS bands. We therefore perform a cross-matching of the Alksnis's catalogue with the 2MASS PSC. Then, in order to derive the absolute magnitude, we need to get an accurate estimation of the distance to each star in that catalogue. 

The recent \textit{Gaia} DR2 gives us the parallax to an unprecedented number of stars in our Galaxy. However, for the vast majority of stars in the second \textit{Gaia} data release, reliable distances cannot be obtained by inverting the parallax. Indeed, an accurate inference procedure must instead be used to account for the nonlinearity of the transformation and the asymmetry of the resulting probability distribution. Consequently, we use the catalogue by \citet{2018AJ....156...58B}, in which the authors derived purely geometric distance estimates using \textit{Gaia} DR2. The authors use a weak distance prior that varies smoothly as a function of galactic longitude and latitude according to a galaxy model. Therefore, these distances are valid as long as the assumed structure of the MW is valid.

After these two cross-matchings we obtain a catalogue of carbon stars containing near-infrared photometry and distance estimates for each star. As for the LMC and the SMC, we now keep only the sources with a five-parameter solution and we discard spurious solutions. We then obtain colour excess values for this catalogue. Since these stars are in our Galaxy, we need colour-excess values that take into account the distance to the star. We use the recent 3D dust map of the MW, made with Pan-STARRS 1, by \citet{2019ApJ...887...93G, 2018MNRAS.478..651G}. This map gives the quantity of dust reddening in an arbitrary unit; we need to use the extinction coefficients from table~\ref{table:reddeningMW}, in order to get a physically meaningful extinction in the 2MASS bands. After these cross-correlations we retrieved $3744$ stars from the $6891$ stars of the Alksnis catalogue.

The absolute magnitude in 2MASS bands is then given by:
\begin{equation}
    M_x = J - 5\log{d} + 5  - R_x E_\text{Pan},
    \label{equation:absolutemagnitudeMW}
\end{equation}
where $x \in \{J, H, K_s\}$, $R_x$ is the extinction coefficient given in table \ref{table:reddeningMW}, $d$ is the estimated distance from the Bailer-Jones catalogue, and $E_\text{Pan}$ is the reddening from the 3D dust map.

\begin{table}
\caption{Reddening coefficients in 2MASS bands for the MC \citep{2019ApJ...887...93G}}
\centering
\begin{tabular}{l c c c c}
\hline\hline
& $R_J$ & $R_H$ & $R_{K_s}$\\ [0.5ex]
\hline
\textbf{MW} & 0.7927 & 0.4690 & 0.3026 \\[1ex]
\hline
\hline
\end{tabular}
\label{table:reddeningMW}
\end{table}

Similarly the intrinsic colours are given by 
\begin{equation}
\begin{split}
    (J-K_s)_0 =~& (J-K_s) - (R_J -  R_{K_s})E_\text{Pan}\\
    (J-H)_0 =~& (J-H) - (R_J -  R_H)E_\text{Pan}\\
    (H-K_s)_0 =~& (H-K_s) - (R_H -  R_{K_s})E_\text{Pan},
    \end{split}
    \label{equation:colorsMW}
\end{equation}
where $J-K_s$, $J-H$ and $H-K_S$ are the observed colour from the 2MASS PSC.

\begin{figure}
\includegraphics[width=\columnwidth, keepaspectratio=true]{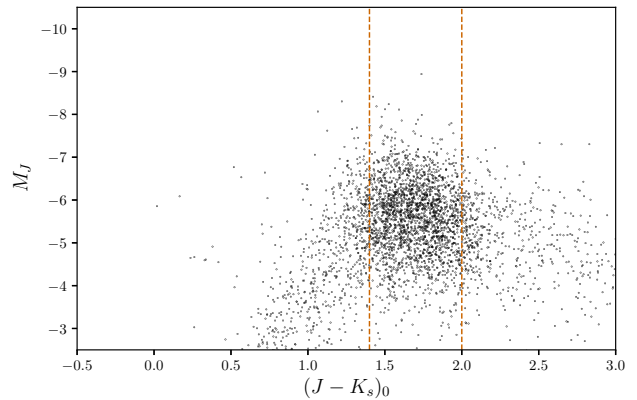}
\caption{Colour-magnitude diagram of the carbon stars in the Milky Way (from the Alksnis catalogue).}
\label{fig:CMDMW}
\end{figure}

We plot the CMD of the galactic carbon stars in Fig.~\ref{fig:CMDMW}. The carbon-star feature is much wider in magnitude than in the MC. The Alksnis's catalogue also contains warmer carbon stars (C-R or C-H types).

We perform the same colour selection as in the LMC and SMC (see Fig.~\ref{fig:CMDMW}).

\subsection{Spectroscopic catalogue}
\label{section:spectroscopic_MW}
\begin{figure}
\includegraphics[width=\columnwidth, keepaspectratio=true]{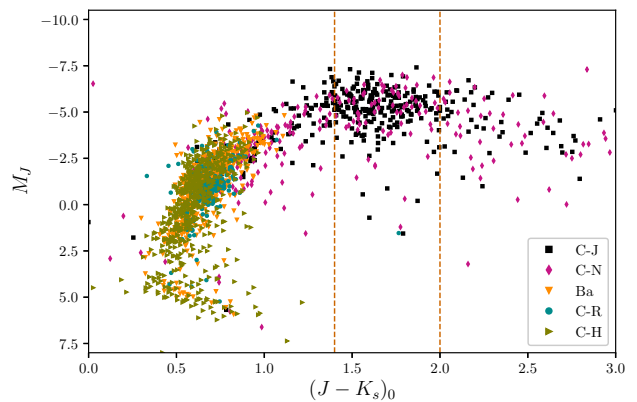}
\caption{Colour-magnitude diagram of the carbon stars in the Milky Way, from the Li catalogue. The different spectral types follow the revised Morgan–Keenan classification.}
\label{fig:CMDLi}
\end{figure}

The spectroscopic catalogue of carbon-rich stars in our Galaxy by \citet{2018ApJS..234...31L} contains the spectral type of 2651 carbon stars from the fourth Data Release (DR4) of the Large Sky Area Multi-Object Fiber Spectroscopy Telescope (LAMOST). We reduced this catalogue the same way as for the Alksnis catalogue and plotted the CMD of these stars in Fig.~\ref{fig:CMDLi}. We clearly see that the bright carbon stars falling in our colour selection are essentially of types C-N and C-J (following the revised Morgan–Keenan classification).

\subsection{Carbon-star luminosity function in the Milky Way}
\label{section:CSLFMW}

We derived the CSLF of the MW for all the stars within the colour range: 
\begin{equation}
    1.4 < (J-K_s)_0 < 2.
\end{equation}

Although Gaia parallaxes can be trusted \citep{2018A&A...616A...9L}, we find that most of our stars have a distance error (converted to absolute magnitude) of same order as the dispersion the carbon stars in the CMD. Therefore, we use maximum likelihood statistics in order to derive the intrinsic distribution of carbon stars in the Milky Way. We use the same assumptions as for the MC. We obtain the following statistical parameters:

\begin{equation} 
\left\{
    \begin{array}{l}
    \bar{M_J} = -5.601 \pm 0.026\\
    \sigma = 0.674 \pm 0.019
    \end{array}
\right.
\end{equation}

The distribution of carbon stars in the Milky Way is much wider than in the MC. We believe that this effect is due to the uncertainty in the distance to each star. Better distance estimations would probably narrow down the distribution to one similar to the MC.

\section{Discussion}
\label{section:discussionMC}
\subsection{A comparison of the carbon-star luminosity functions}
\begin{figure}
\includegraphics[width=\columnwidth]{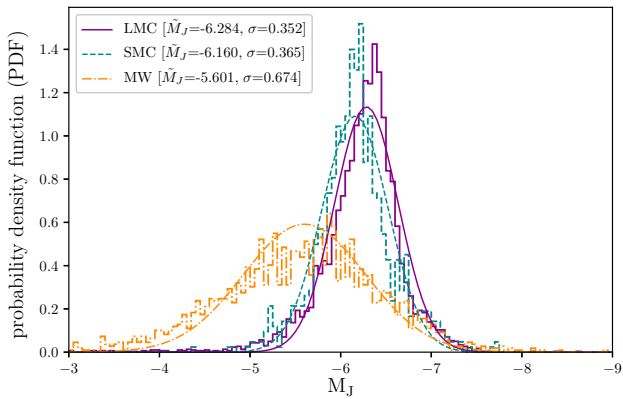}
\caption{Carbon-star luminosity functions in the Large Magellanic Cloud (solid line, dark magenta; 6009 stars), Small Magellanic Cloud (dashed line, dark cyan; 843 stars) and the Milky Way (dot-dashed line, dark orange; 2185 stars). Histograms depict 2MASS data; curves represent the associated normal distributions .}
\label{fig:CSLFcomparisonMW}
\end{figure}

We now compare the CSLFs of both of the MC and the MW (see Fig.~\ref{fig:CSLFcomparisonMW}).

\subsubsection{The Magellanic Clouds}
According to \citet{1999A&A...344..123M}, the carbon-star distribution is mainly determined by the third-dredge-up properties (mass loss and envelope burning); star formation rate (SFR) and initial mass function have a much weaker effect. \citet{1999A&A...344..123M} studied the third-dredge-up parameters in the MC, for metallicites $Z \leq 0.008$. The authors found that the fainter peak of the SMC requires a greater efficiency of dredge-up. This is in agreement with the lower metallicity of the SMC, since at lower metallicities fewer dredge-up episodes are required to convert an oxygen-rich star into a carbon star, due to the smaller initial amount of oxygen in the envelope \citep{1983ApJ...264..114R, 1999A&A...344..123M, 1999ASPC..165..264R, 2003MNRAS.338..572M,2017ApJ...835...77M}. They also found that the SMC distribution is expected to be broader. In addition, \citet{1981ApJ...249..481C, 2003A&A...406...51C} claim that the $(J-K_s)_0$ colour that discriminates carbon stars from oxygen-rich stars decreases with metallicity. In other words, SMC carbon stars should be bluer than the LMC ones.

We find the median J-band absolute magnitude of the CSLF in the SMC to be fainter by $0.124 \pm 0.016$ mag than in the LMC. Moreover, carbon stars in the SMC seem to be bluer by about 0.1 magnitude. Finally, both CSLFs have approximately the same dispersion; it only differs by $0.013 \pm 0.015$ mag, which is of same order of magnitude of the reddening uncertainty in the MC. Our results are in agreement with the claim that the differences between the CSLFs in the LMC and SMC are due to a different metallicity.

Furthermore, we notice that the two CSLFs also have different shapes, with a drop at high luminosities for the CSLF in the LMC. According to \citet{1999A&A...344..123M}, while the faint end depends on dredge-up properties (essentially the temperature parameter), the bright end is sensitive to the recent history of star formation. They found that the CSLF in the SMC is well reproduced with a constant SFR. However, in the LMC, a recent drop of the SFR could explain the observed slight lack of bright carbon stars. 

\subsubsection{The Milky Way}
On the other hand, although stars were selected the same way as in the MC, the distribution of carbon stars in the Milky Way is significantly different. Indeed, the distribution is much broader and its peak is considerably fainter than in the MC.

The MW is known to be more metal-rich than the MC. Therefore one would expect a brighter peak and a narrower distribution, if following the arguments cited above. The only agreement with a higher metallicity is that MW carbon stars appear to be redder than MC carbon stars, as one can clearly see by comparing the different CMDs.

The fact that the properties of the observed CSLF differ from the expected behaviour at higher metallicites could be due to the uncertainty in the distance estimations. Indeed, galactic carbon stars were each assigned a distance which was derived from the \textit{Gaia}-DR2 data. As we mentioned before, the distances induce a median uncertainty of $0.5$ mag, which is of same order as the measured dispersion of galactic carbon stars. Consequently, a better estimation of galactic distances is needed for a better understanding of the distribution of galactic carbon stars. 

Furthermore, differences in molecular gas features and dust emission of carbon stars between the MC and the MW have also recently been found by \citet{2019ApJ...887...82K}. This could play a role in making galactic carbon stars dimmer in the near-infrared. A full study of galactic-carbon-star properties is however beyond the scope of this paper. See \citet{2020A&A...633A.135A} for a study of carbon stars in the solar neighbourhood; one should note that they find the same order of magnitude as ours, for the dispersion of the distribution.

\subsection{Accounting for metallicity effects in the MC}
It is clear that metallicity plays an important role in carbon-star distributions in the Magellanic Clouds. \citet{2009ApJ...690..389M} address metallicity systematics in the luminosity of the TRGB. Following their modified detection method, we use the following composite magnitude that we name C here:
\begin{equation}
    C = M_J - \alpha(J-K_s)_0 + \alpha \gamma,
    \label{eq:linear_model}
\end{equation}
where $\alpha$ is the slope of the carbon-star feature (in J band) for a given galaxy, and $\gamma$ is the metallicity correction. $C$ is the colour-corrected $J$-band magnitude constructed to be independent of and insensitive to metallicity.

In order to find the parameter $\alpha$, we fit a linear regression model to the carbon-star feature for carbon stars in the colour window mentioned above:
\begin{equation}
    M_J = \alpha(J-K_s)_0 - \beta.
\end{equation}
We find the slope $\alpha$ and the intercept $\beta$, for the LMC and the SMC:
\begin{equation} 
\left\{
    \begin{array}{l}
    \alpha_{\text{LMC}} = -0.039 \pm 0.038\\
    \beta_{\text{LMC}} = -6.190 \pm 0.062 \\
    \alpha_{\text{SMC}} = 0.100 \pm 0.098 \\
    \beta_{\text{SMC}} = -6.346 \pm 0.159.
    \end{array}
\right.
\end{equation}
The slopes have errors of almost 100 \% of their values, this means that the features are almost flat in $J$ band, in the MC. The metallicity correction is given by the intersection of these two linear regressions, we obtain:
\begin{equation}
     \gamma = 1.12. 
\end{equation}
This choice of $\gamma$ ensures that the error-weighted mean C magnitude of the two populations are equal.

\begin{figure}
\includegraphics[width=\columnwidth]{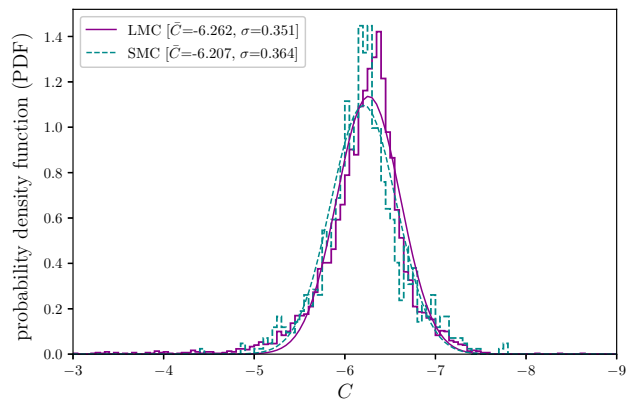}
\caption{Carbon-star luminosity functions in the Large Magellanic Cloud (solid line, dark magenta; 6009 stars) and Small Magellanic Cloud (dashed line, dark cyan; 843 stars), in the C band. Histograms depict 2MASS data; curves represent the associated normal distributions.}
\label{fig:CSLFcomparisonCband}
\end{figure}

We then use maximum likelihood statistics in order to derive the intrinsic distribution of carbon stars in the C band, for the MC. Using Eq.~\ref{eq:maximumlikelihood}, we get for median magnitude and dispersion of the LMC:
\begin{equation} 
\left\{
    \begin{array}{l}
    \bar{C} = -6.262 \pm 0.022\\
    \sigma = 0.351 \pm 0.005, 
    \end{array}
\right.
\end{equation}
and of the SMC:
\begin{equation} 
\left\{
    \begin{array}{l}
    \bar{C} = -6.207 \pm 0.049\\
    \sigma = 0.364 \pm 0.014.
    \end{array}
\right.
\end{equation}
The median C magnitude of the SMC is fainter than the LMC by only $0.055 \pm 0.054$ mag, and the sigmas agree within their error bars. We plotted the luminosity functions in the C band for the Magellanic Clouds in Fig.~\ref{fig:CSLFcomparisonCband}. This method will reduce the impact of metallicity on the true modulus of distant galaxies as well as its final uncertainty.

\section{Conclusions}

\label{section:conclusion}

The Magellanic Clouds, two irregular dwarf galaxies of lower metallicity than the Milky Way, host large numbers of luminous AGB carbon stars. Carbon stars in these two galaxies appear as a distinct horizontal feature in the near-infrared colour-magnitude diagram (absolute magnitude in the J band of 2MASS). They are redder than oxygen-rich AGB stars and also lie in a specific place of the near-infrared colour-magnitude diagram. 

We have developed a method, based purely on two near-infrared bands, to identify carbon stars in the Magellanic Clouds. This method relies on an accurate measurement of the distance to the LMC (first rung of the cosmic distance ladder), and the SMC. Our method uses only on a selection within a $(J-K_s)_0$ colour range, but is affected by metallicity.

As we will see as we explore more distant galaxies, the CSLF is markedly different for grand design spirals (such as the MW) than it is for actively star-forming Magellanic-type irregular galaxies. For this reason when we determine the distances to galaxies in this program, we will restrict ourselves to galaxies similar to the Magellanic Clouds.

Carbon stars are brighter than the TRGB. They are also more luminous than Cepheids and require only a single photometric measurement to establish their luminosity. Consequently, thanks to the next generation of telescopes (\textit{JWST}, \textit{ELT}, \textit{TMT}), carbon stars could be detected in MC-type galaxies at distances of about $50$-$60$ Mpc. Those distances are large enough that the effects of the galaxies' peculiar velocities are not important, yielding a two-step distance ladder to the scale of the Universe. The final goal is to eventually improve the measurement of the Hubble constant and explore the current tensions surrounding its value.

\section*{Acknowledgements}
This work was supported by the Natural Sciences and Engineering Research Council of Canada, the Canada Foundation for Innovation, the British Columbia Knowledge Development Fund.

This publication makes use of data products from the Two Micron All Sky Survey, which is a joint project of the University of Massachusetts and the Infrared Processing and Analysis Center/California Institute of Technology, funded by the National Aeronautics and Space Administration and the National Science Foundation.

This work has made use of data from the European Space Agency (ESA) mission \textit{Gaia} (\url{https://www.cosmos.esa.int/gaia}), processed by the \textit{Gaia} Data Processing and Analysis Consortium (DPAC, \url{https://www.cosmos.esa.int/web/gaia/dpac/consortium}). Funding for the DPAC has been provided by national institutions, in particular the institutions participating in the \textit{Gaia} Multilateral Agreement.

This research has made use of National Aeronautics and Space Administration's Astrophysics Data System Bibliographic Services. Finally, we thank the anonymous referee for carefully reading the manuscript and providing valuable comments.

The codes and catalogues developed and used throughout this work are available at: \url{https://gitlab.com/pripoche/using-carbon-stars-as-standard-candles}.




\bibliographystyle{mnras}
\bibliography{bibliography}


\bsp	
\label{lastpage}
\end{document}